\input harvmac
\input epsf
\input amssym
%
\noblackbox
\newcount\figno
\figno=0
\def\fig#1#2#3{
\par\begingroup\parindent=0pt\leftskip=1cm\rightskip=1cm\parindent=0pt
\baselineskip=11pt
\global\advance\figno by 1
\midinsert
\epsfxsize=#3
\centerline{\epsfbox{#2}}
\vskip -21pt
{\bf Fig.\ \the\figno: } #1\par
\endinsert\endgroup\par
}
\def\figlabel#1{\xdef#1{\the\figno}}
\def\encadremath#1{\vbox{\hrule\hbox{\vrule\kern8pt\vbox{\kern8pt
\hbox{$\displaystyle #1$}\kern8pt}
\kern8pt\vrule}\hrule}}

\def\frac#1#2{{#1 \over #2}}

\def\p{\partial}
\def\semi{\subset\kern-1em\times\;}
\def\bar#1{\overline{#1}}
\def\sqr#1#2{{\vcenter{\vbox{\hrule height.#2pt
\hbox{\vrule width.#2pt height#1pt \kern#1pt \vrule width.#2pt}
\hrule height.#2pt}}}}

\def\p{\partial}

\def\th{\theta}

\def\Db{\bar{D}}

\def\ap{\alpha'}

\def\tb{\overline{\theta}}

\def\psit{\tilde{\psi}}
\def\phit{\tilde{\phi}}

\def\thetat{\tilde{\theta}}

\def\thetah{\hat{\theta}}
\def\phih{\hat{\phi}}
\def\psih{\hat{\psi}}
\def\rh{\hat{r}}
\def\th{\hat{t}}

\def\tt{\tilde{t}}
\def\rt{\tilde{r}}

\def\thetab{\overline{\theta}}
\def\Rt{\tilde{R}}

\def\p{\partial}

%
\def\ap{\alpha'}

\def\tb{\overline{\theta}}

\def\psit{\tilde{\psi}}
\def\phit{\tilde{\phi}}

\def\ZZ{\Bbb{Z}}

\def\IR{\Bbb{R}}


\lref\AdSBH{
J.~B.~Gutowski and H.~S.~Reall,
  ``General supersymmetric AdS(5) black holes,''
  JHEP {\bf 0404}, 048 (2004)
  [arXiv:hep-th/0401129].
}
\lref\myers{ R.~C.~Myers,
``Dielectric-branes,''
JHEP {\bf 9912}, 022 (1999)
[arXiv:hep-th/9910053].
 }
\lref\supertube{
D.~Mateos and P.~K.~Townsend,
``Supertubes,''
Phys.\ Rev.\ Lett.\  {\bf 87}, 011602 (2001)
[arXiv:hep-th/0103030].
}

\lref\mathurlunin{
O.~Lunin and S.~D.~Mathur,
``Metric of the multiply wound rotating string,''
Nucl.\ Phys.\ B {\bf 610}, 49 (2001)
[arXiv:hep-th/0105136].
}

\lref\LuninJY{
O.~Lunin and S.~D.~Mathur,
``AdS/CFT duality and the black hole information paradox,''
Nucl.\ Phys.\ B {\bf 623}, 342 (2002)
[arXiv:hep-th/0109154].
}

\lref\mathurstretch{
O.~Lunin and S.~D.~Mathur,
 ``Statistical interpretation of Bekenstein entropy for systems with a
stretched horizon,''
Phys.\ Rev.\ Lett.\  {\bf 88}, 211303 (2002)
[arXiv:hep-th/0202072].
}

\lref\LuninBJ{
O.~Lunin, S.~D.~Mathur and A.~Saxena,
``What is the gravity dual of a chiral primary?,''
Nucl.\ Phys.\ B {\bf 655}, 185 (2003)
[arXiv:hep-th/0211292].
}

\lref\lmm{
  O.~Lunin, J.~Maldacena and L.~Maoz,
  ``Gravity solutions for the D1-D5 system with angular momentum,''
  arXiv:hep-th/0212210.
}

\lref\mathur{
S.~D.~Mathur, A.~Saxena and Y.~K.~Srivastava,
``Constructing 'hair' for the three charge hole,''
arXiv:hep-th/0311092.
}
\lref\MathurSV{
S.~D.~Mathur,
``Where are the states of a black hole?,''
arXiv:hep-th/0401115.
}

\lref\bmpv{
J.~C.~Breckenridge, R.~C.~Myers, A.~W.~Peet and C.~Vafa,
``D-branes and spinning black holes,''
Phys.\ Lett.\ B {\bf 391}, 93 (1997)
[arXiv:hep-th/9602065].
}

\lref\emparan{
R.~Emparan, D.~Mateos and P.~K.~Townsend,
``Supergravity supertubes,''
JHEP {\bf 0107}, 011 (2001)
[arXiv:hep-th/0106012].
}

\lref\MateosPR{
D.~Mateos, S.~Ng and P.~K.~Townsend,
``Tachyons, supertubes and brane/anti-brane systems,''
JHEP {\bf 0203}, 016 (2002)
[arXiv:hep-th/0112054].
}

\lref\Buscher{T.~H.~Buscher, Phys.\ Lett.\ B {\bf 159}, 127
(1985),\ B {\bf 194}, 59 (1987),\ B  {\bf 201}, 466 (1988),
}

\lref\MeessenQM{ P.~Meessen and T.~Ortin, ``An Sl(2,Z) multiplet
of nine-dimensional type II supergravity theories,'' Nucl.\ Phys.\
B {\bf 541}, 195 (1999) [arXiv:hep-th/9806120].
}

\lref\CallanHN{ C.~G.~Callan, J.~M.~Maldacena and A.~W.~Peet,
``Extremal Black Holes As Fundamental Strings,'' Nucl.\ Phys.\ B
{\bf 475}, 645 (1996) [arXiv:hep-th/9510134].
}

\lref\DabholkarNC{ A.~Dabholkar, J.~P.~Gauntlett, J.~A.~Harvey and
D.~Waldram, ``Strings as Solitons \& Black Holes as Strings,''
Nucl.\ Phys.\ B {\bf 474}, 85 (1996) [arXiv:hep-th/9511053].
}

\lref\CveticXH{ M.~Cvetic and F.~Larsen, ``Near horizon geometry
of  rotating black holes in five dimensions,'' Nucl.\ Phys.\ B
{\bf 531}, 239 (1998) [arXiv:hep-th/9805097].
}

\lref\EmparanWN{
R.~Emparan and H.~S.~Reall,
``A rotating black ring in five dimensions,''
Phys.\ Rev.\ Lett.\  {\bf 88}, 101101 (2002)
[arXiv:hep-th/0110260].
}

\lref\ReallBH{
H.~S.~Reall,
``Higher dimensional black holes and supersymmetry,''
Phys.\ Rev.\ D {\bf 68}, 024024 (2003)
[arXiv:hep-th/0211290].
}

\lref\ElvangMJ{
H.~Elvang and R.~Emparan,
 ``Black rings, supertubes, and a stringy resolution of black hole
non-uniqueness,''
JHEP {\bf 0311}, 035 (2003)
[arXiv:hep-th/0310008].
}

\lref\EmparanWY{
R.~Emparan,
``Rotating circular strings, and infinite non-uniqueness of black rings,''
arXiv:hep-th/0402149.
}

\lref\BenaWV{
  I.~Bena,
  ``Splitting hairs of the three charge black hole,''
  Phys.\ Rev.\ D {\bf 70}, 105018 (2004)
  [arXiv:hep-th/0404073].
}

\lref\BalasubramanianRT{ V.~Balasubramanian, J.~de Boer,
E.~Keski-Vakkuri  and S.~F.~Ross, ``Supersymmetric conical
defects: Towards a string theoretic description  of
black hole formation,''
Phys.\ Rev.\ D {\bf 64}, 064011 (2001) [arXiv:hep-th/0011217].
}

\lref\MaldacenaDR{ J.~M.~Maldacena and L.~Maoz,
``De-singularization  by rotation,'' JHEP {\bf 0212}, 055 (2002)
[arXiv:hep-th/0012025].
}

\lref\BW{ I.~Bena and N.~P.~Warner, ``One Ring to Rule Them All
... and  in the Darkness Bind Them?,'' arXiv:hep-th/0408106.
}

\lref\MaharanaMY{
  J.~Maharana and J.~H.~Schwarz,
  ``Noncompact symmetries in string theory,''
  Nucl.\ Phys.\ B {\bf 390}, 3 (1993)
  [arXiv:hep-th/9207016].
}

\lref\GauntlettQY{ J.~P.~Gauntlett and J.~B.~Gutowski, ``General
Concentric  Black Rings,'' arXiv:hep-th/0408122.
J.~P.~Gauntlett and J.~B.~Gutowski,
``Concentric  black rings,'' arXiv:hep-th/0408010.
}

\lref\ElvangYY{ H.~Elvang, ``A charged rotating black ring,''
Phys.\ Rev. \ D {\bf 68}, 124016 (2003) [arXiv:hep-th/0305247].
}

\lref\StromingerSH{ A.~Strominger and C.~Vafa, ``Microscopic
Origin  of the Bekenstein-Hawking Entropy,'' Phys.\ Lett.\ B {\bf
379}, 99 (1996) [arXiv:hep-th/9601029].
}

\lref\ConstableDJ{ N.~R.~Constable, C.~V.~Johnson and R.~C.~Myers,
``Fractional branes and the entropy of 4D black holes,'' JHEP {\bf
0009}, 039 (2000) [arXiv:hep-th/0008226].
}

\lref\JohnsonGA{ C.~V.~Johnson, R.~R.~Khuri and R.~C.~Myers,
``Entropy of 4D Extremal Black Holes,''
Phys.\ Lett.\ B {\bf 378}, 78 (1996) [arXiv:hep-th/9603061].
}

\lref\MaldacenaGB{ J.~M.~Maldacena and A.~Strominger,
``Statistical  Entropy of Four-Dimensional Extremal Black Holes,''
Phys.\ Rev.\ Lett.\  {\bf 77}, 428 (1996) [arXiv:hep-th/9603060].
}

\lref\CveticXH{ M.~Cvetic and F.~Larsen, ``Near horizon geometry
of  rotating black holes in five dimensions,'' Nucl.\ Phys.\ B
{\bf 531}, 239 (1998) [arXiv:hep-th/9805097].
}

\lref\GiustoID{ S.~Giusto, S.~D.~Mathur and A.~Saxena, ``Dual
geometries for a set of 3-charge microstates,''
arXiv:hep-th/0405017.
}

\lref\GiustoIP{ S.~Giusto, S.~D.~Mathur and A.~Saxena, ``3-charge
geometries and their CFT duals,'' arXiv:hep-th/0406103.
}

\lref\LarsenUK{ F.~Larsen and E.~J.~Martinec, ``U(1) charges and
moduli in the D1-D5 system,'' JHEP {\bf 9906}, 019 (1999)
[arXiv:hep-th/9905064].
}

\lref\SeibergXZ{ N.~Seiberg and E.~Witten, ``The D1/D5 system and
singular CFT,'' JHEP {\bf 9904}, 017 (1999)
[arXiv:hep-th/9903224].
}

\lref\LuninUU{ O.~Lunin, ``Adding momentum to D1-D5 system,'' JHEP
{\bf 0404}, 054 (2004) [arXiv:hep-th/0404006].
}

\lref\KalloshUY{ R.~Kallosh and B.~Kol, ``E(7) Symmetric Area of
the Black Hole Horizon,'' Phys.\ Rev.\ D {\bf 53}, 5344 (1996)
[arXiv:hep-th/9602014].
}

\lref\BertoliniYA{ M.~Bertolini and M.~Trigiante, ``Microscopic
entropy of the most general four-dimensional BPS black  hole,''
JHEP {\bf 0010}, 002 (2000) [arXiv:hep-th/0008201].
}
\lref\BertoliniEI{ M.~Bertolini and M.~Trigiante, ``Regular BPS
black  holes: Macroscopic and microscopic description of the
generating solution,''
Nucl.\ Phys.\ B {\bf 582}, 393 (2000) [arXiv:hep-th/0002191].
}
\lref\MaldacenaDE{ J.~M.~Maldacena, A.~Strominger and E.~Witten,
``Black hole entropy in M-theory,'' JHEP {\bf 9712}, 002 (1997)
[arXiv:hep-th/9711053].
}

\lref\DavidWN{ J.~R.~David, G.~Mandal and S.~R.~Wadia,
``Microscopic  formulation of black holes in string theory,''
Phys.\ Rept.\ {\bf 369}, 549 (2002) [arXiv:hep-th/0203048].
}

\lref\AharonyTI{ O.~Aharony, S.~S.~Gubser, J.~M.~Maldacena,
H.~Ooguri and  Y.~Oz, ``Large N field theories, string theory and
gravity,'' Phys.\ Rept.\  {\bf 323}, 183 (2000)
[arXiv:hep-th/9905111].
}

\lref\BanadosWN{ M.~Banados, C.~Teitelboim and J.~Zanelli, ``The
Black hole  in three-dimensional space-time,'' Phys.\ Rev.\ Lett.\
{\bf 69}, 1849 (1992) [arXiv:hep-th/9204099].
}

\lref\BrownNW{ J.~D.~Brown and M.~Henneaux, ``Central Charges In
The  Canonical Realization Of Asymptotic Symmetries: An
Example From Three-Dimensional Gravity,''
Commun.\ Math.\ Phys.\  {\bf 104}, 207 (1986).
}

\lref\KutasovZH{ D.~Kutasov, F.~Larsen and R.~G.~Leigh, ``String
theory in magnetic  monopole backgrounds,'' Nucl.\ Phys.\ B {\bf
550}, 183 (1999) [arXiv:hep-th/9812027].
}

\lref\LarsenDH{ F.~Larsen and E.~J.~Martinec, ``Currents and
moduli in the (4,0)  theory,'' JHEP {\bf 9911}, 002 (1999)
[arXiv:hep-th/9909088].
}

\lref\min{ J.~P.~Gauntlett, J.~B.~Gutowski, C.~M.~Hull,
S.~Pakis and H.~S.~Reall,
``All supersymmetric solutions of minimal supergravity in five dimensions,''
Class.\ Quant.\ Grav.\  {\bf 20}, 4587 (2003)
[arXiv:hep-th/0209114].
}

\lref\GutowskiRG{
J.~B.~Gutowski, D.~Martelli and H.~S.~Reall,
``All supersymmetric solutions of minimal supergravity in six
dimensions,'' Class.\ Quant.\ Grav.\  {\bf 20}, 5049 (2003)
[arXiv:hep-th/0306235].
}

\lref\usc{
C.~N.~Gowdigere, D.~Nemeschansky and N.~P.~Warner,
``Supersymmetric solutions with fluxes from algebraic Killing spinors,''
arXiv:hep-th/0306097.
}

\lref\MathurZP{
  S.~D.~Mathur,
  ``The fuzzball proposal for black holes: An elementary review,''
  arXiv:hep-th/0502050.
}

\lref\GiustoKJ{
  S.~Giusto and S.~D.~Mathur,
  ``Geometry of D1-D5-P bound states,''
  arXiv:hep-th/0409067.
}

\lref\GiustoIP{
  S.~Giusto, S.~D.~Mathur and A.~Saxena,
  ``3-charge geometries and their CFT duals,''
  Nucl.\ Phys.\ B {\bf 710}, 425 (2005)
  [arXiv:hep-th/0406103].
}

\lref\GiustoID{
  S.~Giusto, S.~D.~Mathur and A.~Saxena,
  ``Dual geometries for a set of 3-charge microstates,''
  Nucl.\ Phys.\ B {\bf 701}, 357 (2004)
  [arXiv:hep-th/0405017].
}

\lref\LuninUU{
  O.~Lunin,
  ``Adding momentum to D1-D5 system,''
  JHEP {\bf 0404}, 054 (2004)
  [arXiv:hep-th/0404006].
}

\lref\BenaTK{
  I.~Bena and P.~Kraus,
  ``Microscopic description of black rings in AdS/CFT,''
  JHEP {\bf 0412}, 070 (2004)
  [arXiv:hep-th/0408186].
}

\lref\CyrierHJ{
  M.~Cyrier, M.~Guica, D.~Mateos and A.~Strominger,
  ``Microscopic entropy of the black ring,''
  arXiv:hep-th/0411187.
}

\lref\BenaWT{
  I.~Bena and P.~Kraus,
  ``Three charge supertubes and black hole hair,''
  Phys.\ Rev.\ D {\bf 70}, 046003 (2004)
  [arXiv:hep-th/0402144].
}

\lref\ElvangDS{
  H.~Elvang, R.~Emparan, D.~Mateos and H.~S.~Reall,
  ``Supersymmetric black rings and three-charge supertubes,''
  Phys.\ Rev.\ D {\bf 71}, 024033 (2005)
  [arXiv:hep-th/0408120].
}

\lref\ElvangRT{
  H.~Elvang, R.~Emparan, D.~Mateos and H.~S.~Reall,
  ``A supersymmetric black ring,''
  Phys.\ Rev.\ Lett.\  {\bf 93}, 211302 (2004)
  [arXiv:hep-th/0407065].
}

\lref\BenaTD{
  I.~Bena, C.~W.~Wang and N.~P.~Warner,
  ``Black rings with varying charge density,''
  arXiv:hep-th/0411072.
}

\lref\BenaDE{
  I.~Bena and N.~P.~Warner,
  ``One ring to rule them all ... and in the darkness bind them?,''
  arXiv:hep-th/0408106.
}

\lref\GrossHB{
  D.~J.~Gross and M.~J.~Perry,
  ``Magnetic Monopoles In Kaluza-Klein Theories,''
  Nucl.\ Phys.\ B {\bf 226}, 29 (1983).
}

\lref\SorkinNS{
  R.~d.~Sorkin,
  ``Kaluza-Klein Monopole,''
  Phys.\ Rev.\ Lett.\  {\bf 51}, 87 (1983).
}

\lref\KS{
I.~R.~Klebanov and M.~J.~Strassler,
``Supergravity and a confining gauge theory: Duality cascades and
$\chi$SB-resolution of naked singularities,''
JHEP {\bf 0008}, 052 (2000)
[arXiv:hep-th/0007191].

I.~R.~Klebanov and A.~A.~Tseytlin,
``Gravity duals of supersymmetric SU(N) x SU(N+M) gauge theories,''
Nucl.\ Phys.\ B {\bf 578}, 123 (2000)
[arXiv:hep-th/0002159].
}

\lref\GiustoXM{
  S.~Giusto and S.~D.~Mathur,
  ``Fuzzball geometries and higher derivative corrections for extremal holes,''
  arXiv:hep-th/0412133.
}

\lref\denefa{
  F.~Denef,
   ``Supergravity flows and D-brane stability,''
  JHEP {\bf 0008}, 050 (2000)
  [arXiv:hep-th/0005049].
}

\lref\denefc{
  B.~Bates and F.~Denef,
   ``Exact solutions for supersymmetric stationary black hole composites,''
  arXiv:hep-th/0304094.
}

\lref\denefb{
  F.~Denef,
   ``Quantum quivers and Hall/hole halos,''
  JHEP {\bf 0210}, 023 (2002)
  [arXiv:hep-th/0206072].
}

\lref\tong{
  D.~Tong,
``NS5-branes, T-duality and worldsheet instantons,''
  JHEP {\bf 0207}, 013 (2002)
  [arXiv:hep-th/0204186].
}
\lref\news{S.~D.~Mathur,
  ``Gravity on AdS(3) and flat connections in the boundary CFT,''
  arXiv:hep-th/0101118.
}
\lref\newj{ J.~R.~David, G.~Mandal, S.~Vaidya and S.~R.~Wadia,
  ``Point mass geometries, spectral flow and AdS(3)-CFT(2) correspondence,''
  Nucl.\ Phys.\ B {\bf 564}, 128 (2000)
  [arXiv:hep-th/9906112].
}

\Title{
  \vbox{\baselineskip12pt \hbox{hep-th/0503053}
  \hbox{UCLA-05-TEP-07}
  \vskip-.5in}
}{\vbox{
  \centerline{Microstates of the D1-D5-KK System}
 }}

\centerline{Iosif Bena and Per Kraus}

\bigskip\medskip
\centerline{ \it Department of Physics and Astronomy,
UCLA, Los Angeles, CA 90095-1547,
USA}

\medskip
\medskip
\medskip
\medskip
\medskip
\medskip
\baselineskip14pt
\noindent

We find supergravity solutions corresponding to all $U(1) \times
U(1)$ invariant chiral primaries of the D1-D5-KK system.  These
solutions  are $1/8$ BPS, carry angular momentum, and are
asymptotically flat in the $3+1$ dimensional sense.  They can be
thought of as representing the ground states of the four
dimensional black hole constructed from the D1-D5-KK-P system.
Demanding the absence of unphysical singularities in our solutions
determines all free parameters, and gives precise agreement with
the quantum numbers expected from the CFT point of view.  The
physical mechanism behind the smoothness of the solutions is that
the D1-branes and D5-branes expand into a KK-monopole supertube in
the transverse space of the original KK-monopole.

\Date{March, 2005}
\baselineskip14pt
\newsec{Introduction}

The D1-D5-KK system (KK = Kaluza-Klein monopole)  is a $1/8$ BPS
configuration in type IIB string theory.  For large brane charges
there is a large microscopic degeneracy of states,
corresponding to an entropy $S= 2\pi \sqrt{N_1 N_5 N_{K}}$.  At
low energies, the system is described by a $1+1$ dimensional
conformal field theory with $(4,0)$ supersymmetry and central
charge $c= 6 N_1 N_5 N_{K}$, and the microscopic states correspond
to the chiral primaries of this CFT.

In this paper we will find the supergravity geometries dual to a
large class of these microstates.  In particular, we will find all
the geometries which preserve a $U(1)\times U(1)$ symmetry.  These
geometries are of interest for a number of reasons:

\item{(1)} Four dimensional BPS black holes with macroscopic event
horizons can be constructed by wrapping the D1-D5-KK system on
$T^6$ and adding momentum along the intersection of the branes
\JohnsonGA. For large $N_P$, the corresponding black hole has a
Bekenstein-Hawking entropy $S= 2\pi \sqrt{N_1 N_5 N_{K} N_P}$,
which can be accounted for microscopically in the CFT
\refs{\JohnsonGA,\ConstableDJ}. Upon setting $P=0$ so as to obtain
the ``ground states'' of the black hole, one finds that the
geometry develops a naked singularity. A related fact is that this
naive $P=0$ limit yields no trace of the microscopic degeneracy
$S= 2\pi \sqrt{N_1 N_5 N_{K}}$ which we know to be present from
CFT considerations. Our new solutions resolve this puzzle since,
at least in the $U(1)\times U(1)$ invariant case, they provide the
correct geometries which replace the singular limit of the black
hole. This part of our story is directly parallel to the story
involving the zero momentum limit
\refs{\BalasubramanianRT,\MaldacenaDR,\LuninJY} of the rotating
D1-D5-P system \refs{\bmpv,\CveticXH}, which has been much
discussed recently (see \MathurZP\ for a review). In that case the
nonsingular geometries are due to the expansion of the D1 and D5
branes into a KK-monopole supertube \lmm; the smoothness of the
KK-monopole ensures the smoothness of the full geometry. Our
solutions will display a more intricate version of the same
phenomenon.

\item{(2)} As argued by Mathur and collaborators
\refs{\LuninJY,\MathurZP}, if it could be shown that the microstates
of the D1-D5-P system are dual to individual bulk
solutions,\foot{Not all of these solutions are expected to be smooth
semi-classical geometries.} this would give a bulk accounting of the
black hole entropy and lead to a solution of the black hole
information paradox.   Some smooth solutions carrying these charges
have indeed been found
\refs{\LuninUU,\GiustoID,\GiustoIP,\GiustoKJ}. After a chain of
dualities, the D1-D5-KK system can be transformed into the D1-D5-P
system, and so our solutions can be thought of as providing dual
versions of these bulk solutions.  A subtlety, which we discuss more
at the end of the paper, is that for this to be seen explicitly one
must go beyond the supergravity approximation in performing the
T-duality along the KK-monopole fibre.  Assuming that this in
principle can be done, and assuming that we can eventually relax the
condition of $U(1)\times U(1$) invariance, it may be possible to
account for all the entropy of the D1-D5-P black hole in this way.
The key point is that we are finding microstates of a genuine
3-charge system, which up to dualities, corresponds to a black hole
with macroscopic event horizon.  See \GiustoXM\ for another recent
discussion of the relationship between these two systems.

\lref\SugawaraQP{
  Y.~Sugawara,
  ``N = (0,4) quiver SCFT(2) and supergravity on AdS(3) x S(2),''
  JHEP {\bf 9906}, 035 (1999)
  [arXiv:hep-th/9903120].
}

\item{(3)}  Studies of three-charge supertubes \refs{\BenaWT,\BenaWV} recently
led to the prediction and subsequent discovery of new BPS black ring
solutions \refs{\ElvangRT,\BW,\ElvangDS,\GauntlettQY,\BenaTD}. In
the type IIB duality frame, these black rings carry the charges of
the D1-D5-P system.  Furthermore, they have a macroscopic entropy
that can be accounted for (modulo some assumptions which remain to
be fully understood) via two microscopic routes.  In one approach
\BenaTK\ (see also \CyrierHJ), one notes that the branes that make
the BPS black ring are the same as the branes that make the 4D black
hole, and so one can map the microscopic computation of the black
ring entropy to that of the 4D black hole discussed above. This
indeed yields agreement with the black ring entropy formula. As
above, it is interesting to consider the limit in which the
macroscopic entropy of the black ring is taken to zero; if smooth
geometries result then these will yield smooth microstates of the
D1-D5-P system. However, this limit yields singular geometries,
which is in fact expected since the geometry near the ring is dual
to the ``zero entropy'' limit of the 4D black hole, whose naive
geometry is singular. The solutions we find resolve the singularity
of the naive zero momentum limit of the D1-D5-KK-P 4D black hole,
and it is likely that they can similarly be thought of as resolving
the singularities of the black rings in this limit.  To show this
explicitly one must ``glue'' our solutions into the BPS black ring
geometry, but we leave that for the future.

\noindent We now turn to a summary of our results.  The CFT of the
D1-D5-KK system is similar in many respects to that of the more
familiar D1-D5 system; see
\refs{\MaldacenaDE,\KutasovZH,\LarsenDH,\SugawaraQP} for discussion.
In particular, at the orbifold point one can think of an effective
string of length $N_1 N_5 N_K$ which can be broken up into any
number of integer length component strings. Each component string
carries $1/2$ unit of four-dimensional angular momentum via fermion
zero modes. Therefore, the microstates carry angular momentum in the
range ${1 \over 2} \leq |J| \leq \half N_1 N_5 N_K$.   The $U(1)
\times U(1)$  invariant microstates whose geometries we find in this
paper correspond to collections of component strings of equal
length.   Our solutions will
 thus be labeled by the number of component strings $n$, and will
carry angular momentum $J = \pm \half {N_1 N_5 N_K \over n}$ with
$1 \leq n \leq N_1 N_5 N_K$.

Our solutions are asymptotically flat in four dimensions, and have
mass $M\propto Q_1 +Q_5 +Q_K$ as follows from BPS considerations.
For $n=1$ the solutions are completely smooth in the
ten-dimensional sense, while for general $n$ they have $\ZZ_n$
singularities caused by the presence of $n$ coincident KK-monopoles; from
the point of view of string theory these are familiar and harmless
orbifold singularities.

\lref\MartinecXQ{
  E.~J.~Martinec and W.~McElgin,
  ``Exciting AdS orbifolds,''
  JHEP {\bf 0210}, 050 (2002)
  [arXiv:hep-th/0206175]; ``String theory on AdS orbifolds,''
  JHEP {\bf 0204}, 029 (2002)
  [arXiv:hep-th/0106171].
}

It will turn out that in addition to carrying D1-D5-KK charge, our
solutions will also carry an electric charge with respect to the
gauge field under which the KK-monopole is magnetically charged.
The solutions will carry $N_e$ units of electric charge subject to
the condition
\eqn\aa{ J = \half N_e N_K~.}
This is the same angular momentum that results in
ordinary electromagnetism from having separated electric and magnetic
charges; the angular momentum is generated by the crossed
electric and magnetic fields. Although we thus have an additional
charge as measured at infinity, this charge disappears after taking
the near horizon limit of our solutions. In this limit, our
solutions reduce to certain BPS conical defect orbifolds of AdS$_3
\times S^3/\ZZ_{N_K}  \times T^4$, closely related to similar
conical defects in the D1-D5 system
\refs{\BalasubramanianRT,\MaldacenaDR,\LuninJY,\MartinecXQ,
\news,\newj}.   Since
these conical defect geometries are known to correspond to the CFT
microstates this confirms that we have indeed constructed
microstates of the D1-D5-KK CFT. In fact, a useful method of
constructing our solutions is to start from the conical defects and
then to try to extend them to the asymptotically flat region,
although we will see that this is much more involved than simply
inserting $1$'s in harmonic functions.

The essential mechanism that renders our solutions smooth is the
expansion of the D1 and D5 branes into a KK-monopole supertube, as
in \lmm.   This is seen in the 10D metric by the fact that one has
harmonic functions sourced on a ring rather than just at a point,
as is the case in the naive singular geometry.  A novel feature in
our case is that we will have harmonic functions sourced both on a
ring and at the origin.   The latter corresponds to the original
KK-monopole.  Thus we have a separation of the D1 and D5 charges
from that of the KK-monopole.  Hence, from a string theory
perspective the singularity is resolved because the D1-D5 system
expands into a supertube in the Taub-NUT geometry of the
KK-monopole. From a 4D perspective  this singularity resolution
does not appear to come from an expansion of the branes (the
supertube formed by the D1 and D5 branes reduces to a point when
compactified to 4D), but from the separation of the branes that
form the D1-D5-KK system into two separated stacks. We believe
that this separation of charges is a more generic phenomenon which
will play a crucial role in providing smooth geometries for other
multi-charge systems.  For instance, it is a basic aspect of the
split attractor flows studied in
 \refs{\denefa,\denefb,\denefc}, as is the angular momentum
 formula \aa.

In Section 2  we use the D1-D5-KK CFT to find the near-horizon limit of and
motivate an {\it ansatz} for the asymptotically flat solutions which are then
constructed in Section 3 and
summarized in Section 4. In Section 5 we explore the properties of
these solutions, and in Section 6 we discuss our results.
Details of the singularity analysis are found in the Appendix.

\newsec{D1-D5-KK CFT and near-horizon geometries}

\subsec{Naive geometry of the D1-D5-KK system}

The naive geometry of the D1-D5-KK system is obtained by
assembling the three individual brane solutions according to the
harmonic function rule \JohnsonGA.   We first review the
KK-monopole metric by itself, since it plays a distinguished role
in our construction, and is perhaps slightly less familiar than
the D-brane metrics.

The KK-monopole   is obtained by replacing four spatial dimensions
by the Euclidean Taub-NUT metric \refs{\SorkinNS,\GrossHB}:
\eqn\ba{\eqalign {ds^2 &= -dt^2 + \sum_{i=5}^9 dx_i^2 + ds_K^2 \cr
ds_K^2 & =Z_K(dr^2 + r^2 d\theta^2 + r^2 \sin^2 \theta d\phi^2)+{1
\over Z_K}\left(R_K d\psi + Q_K (1+\cos \theta) d\phi\right)^2}}
with
\eqn\bb{ Q_K = \half N_K R_K, \quad Z_K = 1 + {Q_K \over r}~.}
The angular coordinates have the identifications $(\psi,\phi) \cong
(\psi+2\pi,\phi) \cong (\psi,\phi+2\pi)$.   The $\psi$ circle
stabilizes at large $r$ at size $2\pi R_K$, and so the Taub-NUT
metric is asymptotically $\IR^3 \times S^1$.   The KK-gauge field
obtained from reduction on the $S^1$ is equal to $A = -Q_K(1 + \cos
\theta) d\phi$.  This is singular at $\cos \theta =1$ where the
$\phi$ coordinate breaks down. However,  this is just a coordinate
singularity, as it is removed by the shift $\psi \rightarrow \psi
-N_K\phi$.  This shift preserves the coordinate identifications, and
it is this requirement which underlies the quantization of the
magnetic charge in \bb\ with $N_K$ an integer (which we'll take to
be positive).

From now on, we will find it convenient to take the KK-gauge field
to be $A= -Q_K \cos \theta d\phi$ to simplify some algebra. With
this choice of gauge, the angular identifications are $(\psi,\phi)
\cong (\psi+2\pi,\phi) \cong (\psi+N_K\pi, \phi+2\pi)$.

At $r=0$ the $\psi$ circle shrinks to zero size.  For $N_K=1$ it
does so smoothly, and in fact the Taub-NUT metric for $N_K=1$ is
completely smooth.  However, for general $N_K$ there is a
$\ZZ_{N_K}$ singularity at $r=0$.

The harmonic function rule yields the D1-D5-KK metric as
\eqn\bc{ ds^2 = {1 \over \sqrt{Z_1 Z_5}} (-dt^2 + dx_5^2)+
\sqrt{Z_1 Z_5} ds_K^2 + \sqrt{{Z_1 \over Z_5}}ds^2_{T^4}}
with
\eqn\bd{ Z_{1,5} = 1+{Q_{1,5} \over r}~.}
$ds_{T^4}^2$ describes a four-torus of volume $V_4$, and $x_5$ is
periodic: $x_5 \cong x_5 +2 \pi R_5$.   The quantization
conditions on the charges are
\eqn\bez{Q_1 = {(2\pi)^4 g \ap^3 N_1 \over 2 R_K V_4}, 
\quad Q_5 = {g\ap N_5 \over 2 R_K}~. }
The D1-branes are wrapped on the $x_5$ circle and smeared on
$T^4$, while the D5-branes are wrapped on both spaces.  Both
branes are also smeared along the KK-direction $\psi$.  The solution
also has a nontrivial dilaton and RR potentials, which we have
suppressed. The solution is $1/8$ BPS, and the BPS mass formula is
\eqn\bfz{ M = {Q_1 + Q_5 +Q_K\over 4 G_4} ~.}

\subsec{Near horizon limit}

To take the near horizon limit relevant for AdS/CFT we drop the
$1$'s from the harmonic functions $Z_{1,5,K}$.  We also change
coordinates as
\eqn\bg{r ={4 Q_1 Q_5 Q_K \over z^2}, \quad \phi = \phit-\psit,
\quad \psi = \half N_K (\psit+\phit), \quad \theta = 2 \thetat}
which brings the metric to the form
\eqn\bh{ ds^2 = {\ell^2\over z^2}(-dt^2 +dx_5^2+dz^2) + \ell^2
(d\thetat^2 + \sin^2 \thetat d\psit ^2 + \cos^2 \thetat d\phit^2)
+\sqrt{{Q_1 \over Q_5}} ds_{T^4}^2~,}
with
\eqn\bi{ \ell^2 = 4 \sqrt{Q_1 Q_5 Q_K^2}~.}
Since the new angular coordinates have the identifications $
(\psit,\phit) \cong (\psit,\phit+ 2\pi) \cong (\psit +{2\pi \over
N_K},\phit +{2\pi \over N_K})$ we identify the geometry \bh\ as
AdS$_3 \times S^3/\ZZ_{N_K} \times T^4$ \KutasovZH.

For our purposes, a central feature is that the metric \bh\ is
singular at $z=\infty$, since the compact $x_5$ circle shrinks to
zero size.  It is precisely this singularity which our new
solutions will ``resolve".

\subsec{D1-D5-KK CFT}

By the standard reasoning, there is a $1+1$ dimensional CFT dual
to string theory on the background   \bh.  Only a few aspects of
this CFT will be relevant for us.  The central charge is
determined from the Brown-Henneaux formula \BrownNW\ $c = {3 \ell
\over 2 G_3}$.  For the more familiar D1-D5 system this gives
$c=6N_1 N_5$.  As we have seen, the KK-monopoles reduce the volume
of the $S^3$ by a factor of $N_K$, and hence decrease $G_3$ by
this same amount, and so now $c = 6N_1 N_5 N_K$.

The $\ZZ_{N_K}$ identification of the sphere breaks the $SU(2)_L
\times SU(2)_R$ isometry group down to $SU(2)_L$, and this becomes
the R-symmetry of the CFT.  The CFT has a corresponding chiral
$(4,0)$ supersymmetry.  For the asymptotically flat geometries,
the $SU(2)$ R-symmetry will be identified with the four
dimensional angular momentum.

We will be interested in the Ramond ground states of the susy side
of the CFT, or equivalently, the NS sector chiral primaries. These
states are conveniently summarized in the orbifold CFT language,
exactly like in the case of the D1-D5 system.   In particular, one
considers an effective string of length $N_1 N_5 N_K$, which can
be broken up into component strings of integer length.  In the
Ramond vacua, each component string carries $J= \pm \half$, where
$J$ is the diagonal $SU(2)$ generator normalized to have
half-integer eigenvalues.   We therefore find that the complete
system can carry R-charge, or equivalently angular momentum, in
the range
\eqn\ca{ - \half N_1 N_5 N_K \leq J \leq \half N_1 N_5 N_K~.}

A particular subclass of states corresponds to taking all
component strings to have equal length and equal R-charge. These
states are therefore labeled by $n$, the length of component
strings, and their R-charges are
\eqn\cb{ J = \pm \half {N_1 N_5 N_K \over n}, \quad 1\leq n \leq
N_1 N_5 N_K~.}
We will find the  supergravity duals to this class of states. What
makes these states simpler is that their geometries preserve a
$U(1)\times U(1)$ symmetry corresponding to translation in $\psi$
and $\phi$.

\subsec{Spectral flow of near horizon geometry}

As we have seen, the near horizon geometry based on the metric
\bc\ is singular because it yields AdS$_3$ in Poincar\'{e}
coordinates with a compact spatial direction.
Our new geometries will, by contrast, reduce to
 global AdS$_3$ in the near-horizon
limit, and so be free of singularities.\foot{To be precise, the
only singularities will be acceptable $\ZZ_n$ singularities.}

Indeed, we want the near horizon limit of our geometries to be dual
to the Ramond vacua of the CFT, which can be mapped into the
NS-sector chiral primaries by spectral flow. Furthermore, the $U(1)
\times U(1)$ invariant chiral primaries are dual  to conical defect
orbifolds of global AdS$_3$, and in the bulk spectral flow is just a
coordinate transformation. Our strategy is therefore to start from
global AdS$_{3}$ and ``undo'' the spectral flow to obtain the
near-horizon limit of the geometries dual to the Ramond vacua. We
then write these near-horizon solutions in a coordinate system
adapted to the BPS equations, which can be used to extend these
solutions to the asymptotically flat region.

With this in mind, we will now transform the metric of global AdS$_3
\times S^3/\ZZ_{N_K} \times T^4$ into this preferred coordinate
system. This procedure will then suggest a natural {\it ansatz} for
constructing the full asymptotically flat solutions.

We therefore start from
\eqn\da{ ds^2 = - \left( 1 + {\rt^2 \over \ell^2} \right) d\tt^2
+ {d\rt^2 \over
1+ {\rt^2 \over \ell^2}} + \rt^2 d\chi^2 + \ell^2 (d\thetat^2 +
\sin^2 \thetat d\psit ^2 + \cos^2 \thetat d\phit^2)~.}
We have omitted the $T^4$ since it plays no role in what follows.

We now perform the following chain of coordinate transformations:

\item{Step 1:}
\eqn\db{ \chi = {R_K \over \ell^2}x_5, \quad \psit = \psih +{R_K
\over \ell^2}\th, \quad \phit = \phih+ {R_K \over \ell^2}x_5,
\quad \tt = {R_K \over \ell}\th}

\item{Step 2:}
\eqn\dc{ \rho = \sqrt{\rt^2 + R_K^2 \sin^2 \thetat}, \quad \cos
\thetab = {\rt \cos\thetat \over \sqrt{\rt^2 +R_K^2 \sin^2
\thetat}}}

\item{Step 3:}
\eqn\dd{r= {\rho^2 \over Q_K}, \quad \phi = \phih -\psih, \quad
\psi = \half N_K (\psih + \phih), \quad \theta =2 \thetab~,}
with $Q_K$ defined as in \bb.  The final angular coordinates have
periodicities $(\psi,\phi) \cong (\psi+2\pi,\phi) \cong
(\psi,\phi+2\pi)$, and $x_5 \cong x_5 + 2\pi {\ell^2 \over R_K}$.
The metric takes the form
\eqn\de{\eqalign{ 
ds^2&= {1 \over \sqrt{{Z}_1 {Z}_5}}\left[
-(dt+k)^2 +(dx_5 -k -s)^2 \right]+ \sqrt{ Z_1 {Z}_5} ds_{K}^2 \cr 
k & = {\ell^2 \over 4 Q_K} {\Sigma - r -\Rt \over \Sigma} d\psi
-{\ell^2 \over 4R_K } {\Sigma - r -\Rt \over \Sigma} d\phi \cr 
s & = -{\ell^2 \over 2 Q_K}{\Sigma -r \over \Sigma} d\psi - {\ell^2 \over
2R_K }{\Rt \over \Sigma}d\phi \cr 
Z_K &= {Q_{K}\over \Sigma},\quad
{Z}_{1,5} ={Q_{1,5} \over \Sigma} \cr 
\Sigma & = \sqrt{r^2 +\Rt^2 +2
\Rt r \cos \theta}\cr \Rt & = {R_K^2 \over 4 Q_K}~. }}

Note that if we insert $\Rt=0$ and restore the $1$ in $Z_{1,5}$ and
$Z_K$ then we revert back to the metric of \bc.

Since the metric \de\ is, by construction, smooth, our goal is to
extend it to the asymptotically flat region.  In the analogous case
of the D1-D5 system this can be done simply by adding  $1$'s to the
${Z}$ functions.  In our case it turns out to be much more involved.
Therefore, we will just use \de\ as a guide for writing down an
appropriate asymptotically flat {\it ansatz}, but then analyze the
equations of motion independently of the preceding near horizon
construction.  The asymptotically flat metric we eventually find
will, however, turn out to have \de\ as its near horizon limit.

 \newsec{Construction of asymptotically flat solutions}

 For the purposes of writing a supergravity {\it ansatz} it is
 preferable to work in the M-theory frame, where there is more
 symmetry between the types of branes.  However, the IIB duality
 frame is distinguished by the fact that the resulting geometry is
 free of singularities.  This follows from the fact that in this frame
 the branes expand into a KK-monopole supertube, and the
 KK-monopole is a smooth solution.  We will therefore work in the
 IIB frame, so that we can fix all free coefficients by demanding
 smoothness, and be confident that we have constructed a
 legitimate physical solution.

In the M-theory frame, a background that preserves the same
supersymmetries as three sets of M2-branes can be written as
\refs{\BW,\AdSBH}
%
%
%
\eqn\ea{\eqalign{ ds_{11}^2& =  - \left({1 \over Z_1 Z_2
Z_3}\right)^{2/3} (dt+k)^2 + \left( Z_1 Z_2 Z_3\right)^{1/3}
h_{mn}dx^m dx^n \cr &+ \left({Z_2 Z_3 \over
Z_1^2}\right)^{1/3}(dx_1^2+dx_2^2) + \left({Z_1 Z_3 \over
Z_2^2}\right)^{1/3}(dx_3^2+dx_4^2) + \left({Z_1 Z_2 \over
Z_3^2}\right)^{1/3}(dx_5^2+dx_6^2) \cr {\cal A} & = A^1 \wedge dx_1
\wedge dx_2 +A^2 \wedge dx_3 \wedge dx_4 + A^3 \wedge dx_5 \wedge
dx_6}}
where $A^I$ and $k$ are one-forms in the five-dimensional space
transverse to the $T^6$. $h_{mn}$ is a four dimensional hyper-Kahler
metric.

To obtain the solutions in the type IIB frame with D1, D5, and
momentum charges, we KK reduce along $x_6$, and then perform
T-dualities along $x_{3,4,5}$. The three types of M2 branes
become D1 branes, D5 branes and
momentum ($Z_1 \rightarrow Z_5, Z_2 \rightarrow Z_1, Z_3 \rightarrow Z_p $,
and similarly for the $A^I$)
and the resulting string frame background is:
\eqn\ob{\eqalign{ ds_{10}^2& =  - Z_1^{-1/2} Z_5^{-1/2} Z_p^{-1}  (dt+k)^2
+ Z_1^{1/2} Z_5^{1/2} h_{mn}dx^m dx^n \cr
&+ Z_1^{1/2} Z_5^{- 1/2}(dx_1^2 +dx_2^2 + dx_3^2+dx_4^2) +
Z_1^{-1/2} Z_5^{-1/2} Z_p^{-1} (dx_5 + A^p)^2 \cr
e^{2\phi} &= {Z_1 \over Z_5},
\cr F_{(3)} & = (Z_5^{5/4} Z_1^{-3/4} Z_p^{-1/2}) {\star_5}
d A^5 - dA^1 \wedge (dx_5+A^p) }}
where ${\star_5}$ is taken with respect to the
five-dimensional metric that appears in the
first line of \ob.

When written in terms of the ``dipole field strengths''    $\Theta^I$,
\eqn\oc{\Theta^I
\equiv d A^I + d\left(  {dt +k \over Z_I}\right)} the BPS equations
simplify to \refs{\BW,\AdSBH}:
\eqn\of{\eqalign{ \Theta^I  &= \star_4 \Theta^I \cr
\nabla^2  Z^I & = {1 \over 2  } |\epsilon_{IJK}| \star_4 (\Theta^J \wedge \Theta^K) \cr
dk + \star_4 dk &= Z_I \Theta^I~}}
where $\star_4$ is the Hodge dual taken with respect to the
four-dimensional metric $h_{mn}$. We are looking for a solution
describing the dual of microstates of the D1-D5-KK system, so we
take the momentum to zero. This furthermore implies the absence of
dipole charges associated to the momentum charge. Hence $Z_p = 1$,
and $\Theta^1=\Theta^2=0$. Moreover, we are interested in a
solution that has KK-monopole charge, so we take  the transverse
metric $h_{mn} dx^m dx^n = ds_K^2$, the Euclidean Taub-NUT metric of
\ba.

It will also to be convenient to define $s$ as
\eqn\oca{ s \equiv  -A^p - (dt + k)~,}
such that $d s = - \Theta^p $. With these simplifications the metric
is
\eqn\ojb{ ds_{10}^2= {1 \over \sqrt{Z_1 Z_5}}\left[ -(dt+k)^2
+(dx_5 -k -s)^2 \right]+ \sqrt{Z_1 Z_5} ds_{K}^2+\sqrt{{Z_1 \over
Z_5}}ds_{T^4}^2}
where we took $dx_5 \rightarrow dx_5 - dt$ to impose $g_{tt}=-1$
asymptotically. Note that the metric takes the same form as in \de.
The dilaton is
\eqn\ojc{e^{\phi} = \sqrt{ Z_1 \over Z_5}~,}
and the RR fields have an ``electric'' component given by
\eqn\rrfield{C^2_{e} = Z_1^{-1} (dt + k) \wedge (dx^5 - s - k )  }
and a ``magnetic'' component given implicitly by
\eqn\rrfieldm{d C^2_{m} = - \star_4 (d Z_5)   }
where $ \star_4$ is now the Hodge dual on the Taub-NUT metric \ba.

With these definitions, the BPS equations become simply
\eqn\oh{ds=\star_4 ds  = -(dk +\star_4 dk) ,\quad\quad \nabla^2
Z_{1,5} =0~.}
To simplify further, we define
\eqn\ojbb{ a = k + \half s}
so that the full set of equations is:
\eqn\oja{ ds  =\star_4 ds , \quad  da = - \star_4 da, \quad
\nabla^2 Z_1 = \nabla^2 Z_5 =0~.}
Of course, strictly speaking these equations only hold away from
the brane sources which we also need to specify. If we replace the Taub-NUT space
by $\IR^4$, we recover the solutions of \mathurlunin\ and \lmm.

From \oja\ we see that the problem has been reduced to finding
(anti) self-dual 2-forms and harmonic functions on Taub-NUT.  In
fact, we can further reduce the problem of finding the 2-forms to
that of finding harmonic functions, as we now discuss.

\subsec{(anti) self-dual 2-forms and harmonic functions}

As explained above, we need to find closed, (anti) self-dual 2-forms
on the Taub-NUT space.  We are restricting ourselves to $U(1) \times
U(1)$ invariant solutions, where the $U(1)$'s correspond to shifts in
$\psi$ and $\phi$, and so we demand this of our 2-forms and harmonic
functions.

We can approach the problem in the following way.  Write Taub-NUT
in Cartesian coordinates as
\eqn\qa{ds^2 = Z_K d\vec{x}^2 + {1 \over Z_K} (R_Kd\psi  + \vec{A}
\cdot d\vec{x})^2}
with orientation $\epsilon_{\psi 123}>0$.    We have
\eqn\qab{ \epsilon_i^{~jk}\partial_j A_k  = \partial_i Z_K}
where the $i$ indices refer to the flat  metric $d\vec{x}^2$.

Then, any self-dual, closed 2-form $\Theta^+$ takes the form
\eqn\qb{ \Theta^+_{\psi i} = R_K B_i, \quad \Theta^+_{ij}= A_i B_j
- B_i A_j + Z_K \epsilon_{ij}^{~~k} B_k}
where
\eqn\qc{ B_i = \partial_i P^+, \quad \partial_i^2(Z_K P^+) =0}
and $\p_i^2$ is the Laplacian with respect to  $d\vec{x}^2$.

Similarly, any anti-self-dual, closed 2-form $\Theta^-$ takes the
form
\eqn\qbb{ \Theta^-_{\psi i} = R_K B_i, \quad \Theta^-_{ij}= A_i
B_j - B_i A_j - Z_K \epsilon_{ij}^{~~k} B_k}
where
\eqn\qcc{ B_i = \partial_i P^-, \quad \partial_i^2(P^-) =0~.}

In our case, we make the identifications
\eqn\qd{ \Theta^+ = ds, \quad\Theta^- = da~.}
We can specify harmonic functions $Z_K P^+$ and $P^-$, work out
the 2-forms $\Theta^\pm$, and then integrate to find the 1-forms
$s$ and $a$. We have therefore shown that our full solution is
specified by the four harmonic functions $Z_1$, $Z_5$, $Z_K P^+$
and $P^-$.


\newsec{Asymptotically flat solution: results}

Using our previous near horizon solution  \de\ as a guide, we now
look for a nonsingular asymptotically flat solution.  As we have
discussed, the solution is specified by four harmonic functions.
Writing Taub-NUT as in \ba, our requirement of $U(1)\times U(1)$
symmetry means that the harmonic functions should only depend on
$r$ and $\theta$.  It is then easy to check that Laplace's
equation for such functions is the same as on $\IR^3$.  So we just
have to specify locations of our sources, and then our harmonic
functions will be of the form $\sum_i {q_i \over |\vec{x} -
\vec{x}_i|}$.

As in \de\ (but now including the $1$ for asymptotic flatness) we
will take
\eqn\ca{ Z_{1,5} = 1 + {Q_{1,5} \over \Sigma}, 
\quad \Sigma = \sqrt{r^2 +\Rt^2 +2 \Rt \cos \theta}}
corresponding to charges $Q_{1,5}$ placed at a distance $\Rt$
along the negative z-axis.      The charges are quantized as in
\bez.

Next, we need to specify the harmonic functions $P^-$ and $Z_K
P^+$.  A natural ansatz is
\eqn\ra{\eqalign{P^- &= c_1 + {c_2 \over r} + {c_3 \over \Sigma},
\cr Z_K P^+ &=d_1 +{d_2 \over r} + {d_3 \over \Sigma}~.}}

We now need to solve \qd\ to find $s$ and  $a$.  They have the
structure
\eqn\ed{s = s_\psi(r,\theta)d\psi + s_\phi(r,\theta)d\phi, \quad a
= a_\psi(r,\theta)d\psi + a_\phi(r,\theta)d\phi~.}
From \qb-\qbb\ we can immediately read off
\eqn\rb{\eqalign{a_{\psi}& = - R_KP^- =- R_K\left(  c_1 + {c_2
\over r} + {c_3 \over \Sigma} \right) \cr 
s_\psi &= -R_KP^+ +d_5 = -{R_K \over Z_K} \left(d_1 +{d_2 \over r} 
+ {d_3 \over \Sigma}\right)+d_5~.}}

To determine $a_{\phi}$ and $s_\phi$ we solve the second equations
in \qb\ and \qbb, which read
\eqn\cc{\eqalign{ (da)_{ij} &= A_i \p_j P^- - \p_i P^- A_j -Z_K
\epsilon_{ij}^{~~k}\p_k P^- \cr (ds)_{ij} &= A_i \p_j P^+ - \p_i
P^+ A_j +Z_K \epsilon_{ij}^{~~k}\p_k P^+~.}}
Solving the $r\phi$ and $\theta \phi$ components of these
equations yields
\eqn\cd{\eqalign{  a_\phi &  = -P^- Q_K \cos \theta+ c_3 \left(
{Q_K \over
 \Rt} {(r + \Rt \cos \theta) \over \Sigma} - {(r \cos \theta +\Rt) \over \Sigma}
 \right) +(Q_K c_1 -c_2 )\cos \theta + c_4 \cr s_\phi &= -P^+ Q_K \cos \theta  +d_3 {r \cos\theta +\Rt \over \Sigma} + d_2
\cos \theta + d_4~.}}

Finally, from $k = a - \half s$ we have (we omit the $d_5$ term
from $k_\psi$ corresponding to redefining constants)
\eqn\rmm{\eqalign{ k_\psi & =  -R_K(P^- - \half P^+)       \cr
k_\phi & = -(P^- - \half P^+) Q_K \cos \theta  +c_3{Q_K\over
\Rt}{(r+\Rt \cos \theta) \over \Sigma} -(c_3 +\half d_3){(r\cos
\theta +\Rt ) \over \Sigma}\cr & ~~+ (Q_K c_1 -c_2 - \half
d_2)\cos \theta + c_4 -\half d_4~. }}

\subsec{Result of singularity analysis}

We have now specified all quantities appearing in the metric \ojb\
in terms of the constants $c_i$ and $d_i$ and the radius $\Rt$.
Although we have a solution for any choice of constants, we want
to further demand that we have a smooth solution, free of any
singularities.

There are potential singularities at $r=0$ and $\Sigma=0$ where
the harmonic functions diverge.  Furthermore, there are potential
Dirac-Misner string singularities at $\sin\theta =0$ where the
$\phi$ coordinate breaks down.  In Appendix A we analyze all the
conditions for smoothness, and find that all of the coefficients
$c_i$ and $d_i$ are uniquely fixed, along with the ring radius
$\Rt$.  The values obtained are:
\eqn\daa{\eqalign{
c_1 &= {1 \over 2 Q_K} \sqrt{{Q_1 Q_5 \over \tilde{Z}_K}}+{1 \over 2R_K }d_5, 
\quad c_2=0, \quad c_3 ={1 \over 2} \sqrt{{Q_1 Q_5 \over \tilde{Z}_K}}, 
\quad c_4 = -{Q_K \over 2 \Rt} \sqrt{{Q_1 Q_5 \over \tilde{Z}_K}} \cr 
d_1 &= {1 \over  Q_K} \sqrt{{Q_1 Q_5 \over \tilde{Z}_K}}+{1 \over R_K}d_5, 
\quad d_2 = \sqrt{Q_1 Q_5 \tilde{Z}_K}+{Q_K \over R_K}d_5, 
\quad d_3 = -\sqrt{Q_1 Q_5 \tilde{Z}_K}, \quad d_4 =0~,}}
with
\eqn\Db{\tilde{Z}_K = 1+{Q_K \over \Rt}~.}
 The value of $d_5$ is not
determined by the singularity analysis. However, it turns out that
with the constants given by \da\ there ends up being no dependence
on $d_5$ in the solution, so we now set $d_5=0$.

The ring radius $\Rt$ is determined from
\eqn\Dc{ R_5 = {2 \sqrt{Q_1 Q_5 \tilde{Z}_K}\over n}~.}
Here $n$ is any positive integer.  As discussed in the Appendix,
complete smoothness of the geometry requires that we take $n=1$.
Other values of $n$ correspond to allowing $\ZZ_n$ singularities due
to the presence of $n$ coincident KK-monopole supertubes.  These more
general solutions, while singular in supergravity, are nonsingular
from the point of view of string theory.

\subsec{Summary of solution}

Now that we have worked out all the free parameters we can write
down the explicit solution.  For convenience, we collect all the
relevant formulas here.  The type IIB string frame metric,
dilaton, and RR three-form field strength are
\eqn\Dd{\eqalign{ ds_{10}^2&= {1 \over \sqrt{Z_1 Z_5}}\left[
-(dt+k)^2 +(dx_5 -k -s)^2 \right]+ \sqrt{Z_1 Z_5}
ds_{K}^2+\sqrt{{Z_1 \over Z_5}}ds_{T^4}^2 \cr e^\phi &= \sqrt{{Z_1
\over Z_5}} \cr F^{(3)} & = d \left[ Z_1^{-1} (dt + k) \wedge (dx^5
- s - k )\right] - \star_4 (d Z_5)  }}
where $\star_4$ is taken with respect to the metric $ds_K^2$, and
\eqn\Df{ds_K^2  =Z_K(dr^2 + r^2 d\theta^2 + r^2 \sin^2 \theta
d\phi^2)+{1 \over Z_K}\left(R_K d\psi + Q_K \cos \theta d\phi\right)^2}
\eqn\Dg{Z_K = 1 + {Q_K \over r},  \quad Z_{1,5} = 1+ {Q_{1,5} \over
\Sigma}, \quad  \Sigma =\sqrt{r^2 + \Rt^2 +2\Rt r \cos \theta}}
\eqn\Dh{ x_5 \cong x_5 +2 \pi R_5, \quad R_5 = {2\sqrt{ Q_1 Q_5
\tilde{Z}_K} \over n}, \quad \tilde{Z}_K = 1+ {Q_K \over \Rt}~.}

The 1-forms $s$ and $k$ have the structure $s= s_\psi d\psi +
s_\phi d\phi$ (and analogously for $k$) with components
\eqn\dez{\eqalign{
s_\psi & = - {\sqrt{Q_1 Q_5 \tilde{Z}_K}  R_K
\over  Z_K r \Sigma} \left[ \Sigma -r + {r\Sigma \over Q_K
\tilde{Z}_K}\right] \cr 
s_\phi&=-{\sqrt{Q_1 Q_5 \tilde{Z}_K}  \over
\Sigma}\left[\Rt- {\left(\Sigma-{1 \over \tilde{Z}_K}\Sigma
-r\right) \over Z_K}\cos \theta \right]\cr 
k_\psi & = {\sqrt{Q_1 Q_5
\tilde{Z}_K}  R_K Q_K \over 2 \Rt \tilde{Z}_K r Z_K
\Sigma}\left[\Sigma-r-\Rt-{2 \Rt r\over Q_K}\right] \cr 
k_\phi&= -{\sqrt{Q_1 Q_5 \tilde{Z}_K}  Q_K \over 2 \Rt \tilde{Z}_K \Sigma
}\left[ \Sigma -r-\Rt +{(\Sigma -r+\Rt) \over Z_K}\cos \theta\right]~.
}}
The charges $Q_{K}$ and $Q_{1,5}$ are quantized according to \bb\
and \bez.

The free parameters in the solution are the moduli $R_5$, $R_K$,
$V_4$, and $g_s$; the quantized charges $N_K$, $N_1$, and $N_5$; and
the quantized dipole charge $n$. As we explain in the Appendix, it
is also possible to add two constant parameters to $s_\psi$ and
$s_\psi$; after compactifying to four dimensions, one of these
constants corresponds to shifting the modular parameter of the $T^2$
at infinity, and the other is a trivial gauge transformation of one
of the  potentials that are obtained after the reduction.

\newsec{Properties of the solution}

\subsec{Near horizon limit}

As usual, to take the near horizon decoupling limit we take
$\ap\rightarrow 0$, while scaling coordinates and moduli such
that the metric picks up an overall factor of $\ap$. In our case,
this is achieved by the scaling
\eqn\ga{ r\sim (\ap)^{3/2}, \quad \Rt \sim (\ap)^{3/2}, \quad V_4
\sim (\ap)^2, \quad R_K \sim (\ap)^{1/2}~.}
This scaling effectively takes the large charge limit of the
solution, and eliminates for example the $1$ from $Z_{1,5,K}$ and
$\tilde{Z}_K$.  The 1-forms in \dez\ become
\eqn\dea{\eqalign{s_\psi & = - {\sqrt{Q_1 Q_5 \tilde{Z}_K}  R_K
\over Q_K} { \Sigma -r \over \Sigma} \cr s_\phi&=-\sqrt{Q_1 Q_5
\tilde{Z}_K}\Rt  {1 \over \Sigma}\cr k_\psi & = {\sqrt{Q_1 Q_5
\tilde{Z}_K} R_K \over 2 Q_K }{\Sigma-r-\Rt\over \Sigma} \cr k_\phi&
= -{\sqrt{Q_1 Q_5 \tilde{Z}_K} \over 2 }{\Sigma -r-\Rt\over
\Sigma}~.}}

We can now compare with the solution of \de\ obtained by spectral
flow.    First consider the case $n=1$ corresponding to a singly
wound KK-monopole supertube.  We find complete agreement with \de\
after performing the coordinate rescalings
\eqn\gb{ r \rightarrow {R_K^2 \over 4 Q_K \Rt} r, \quad x_5
\rightarrow \sqrt{{R_K^2 \over 4 Q_K \Rt}}x_5~.}
Recalling that \de\ is just a diffeomorphism of \da, this
demonstrates that the near horizon limit of our asymptotically flat
$n=1$ geometries is just AdS$_3 \times S^3/ \ZZ_{N_K} \times T^4$
with AdS$_3$ appearing in global coordinates. In particular, this
makes the smoothness of the near horizon geometry manifest.

%

Now consider the case of general $n$.  We still get the metric \de,
and hence \da, the only difference is that the periodicity in \da\
is
\eqn\gc{ (\chi, \phih)\cong (\chi +{2\pi \over n}, \phih +{2\pi
\over n})~.}
In other words, we have a conical defect.  This has a nice
correspondence with what one expects from the CFT point of view.
In the CFT, states with general $n$ correspond to having component
strings whose length is proportional to $n$.  The energy gap above
the vacuum is therefore of the form  $\Delta E = { \omega_0 \over
n}$.  This is also the case for the conical defect geometries.  To
see this, perform the rescalings
\eqn\gd{ \chi \rightarrow n\chi, \quad r\rightarrow {r \over
n},\quad t \rightarrow {t \over n}}
to bring $\chi$ to the standard $2\pi$ periodicity while
maintaining the asymptotically  global AdS$_3$ form of the metric.
The rescaling of $t$ precisely accounts for the $n$ dependence of
the energy gap.

The fact that our asymptotically flat geometries reduce in the
near horizon limit to geometries with a clear CFT  interpretation
gives us confidence that we have correctly identified our
solutions as the microstates of the D1-D5-KK system.

\subsec{ Smoothness}

We now give a qualitative explanation for the absence of
singularities in our solution.  In the naive solution \bc\ the
D1-branes, D5-branes, and KK-monopoles can all be thought of as
sitting at $r=0$.  The $x_5$ direction common to all the branes
shrinks to zero size at the origin, yielding the singularity. In
the nonsingular solution the D1-branes and D5-branes expand into a
KK-monopole supertube, with the tube direction being the KK fibre
direction of the original KK-monopole.

This is roughly a
supertube in Taub-NUT, to be contrasted with the usual supertube
in $\IR^4$  \refs{\supertube,\emparan,\MateosPR}. From the point of
view of the $\IR^3$ base of the Taub-NUT, the KK-monopole sits at
$r=0$ while the supertube sits at $r=\Rt$ and $\cos \theta=-1$. In
this sense, the KK-monopole charge is separated from the D1-brane
and D5-brane charges.

There is in fact a sort of symmetry between the two types of
charges, as is most readily seen in the context of the near
horizon solution.   In particular, consider the loci $r=0$ and
$\Sigma=0$ corresponding to the ``locations" of the charges. We
ask for their locations in the global AdS metric of \da. Tracing
back through the coordinate transformations \db-\dd, we see that
$r=0$ corresponds to $(\rt=0, \sin \thetat=0)$, and $\Sigma =0$
corresponds to $(\rt=0, \cos \thetat=0)$.   These are identified
as two non-intersecting circles on the $S^3/Z_{N_K}$, centered at
the origin of global AdS$_3$.  So the divergent loci of the two
types of harmonic functions --- $Z_K$ and $Z_{1,5}$ ---  are
simply related by a redefinition of $\theta$.

\subsec{ Kaluza-Klein reduction to four dimensions}

In order to read off the mass, angular momentum, and charge of our
solution it is convenient to reduce it to four dimensions.   This
will also demonstrate that the solution is asymptotically flat in
the four dimensional sense.  The compact directions along which we
reduce are $T^4$ and the asymptotic $T^2$ parameterized by $\psi$
and $x_5$.

First reduce from $D=10$ to $D=6$.  Writing the $D=10$ metric as
\eqn\ua{ds_{10}^2 = ds_6^2 + e^{2\chi} ds_{T^4}^2 }
the $D=6$ metric is $ds_6^2$ and the $D=6$ dilaton is
\eqn\ub{ \phi_6 = \phi_{10} -2\chi = 0~.}

To reduce to $D=4$ we need to write the six dimensional metric as
\eqn\za{\eqalign{ ds_6^2 &= ds_4^2 + G_{\psi \psi} ( R_Kd\psi -
A^{(\psi)}_\mu dx^\mu)^2 +G_{55}( dx_5 - A^{(z)}_\mu dx^\mu)^2 \cr
& \quad+ 2G_{\psi 5} ( R_Kd\psi - A^{(\psi)}_\mu dx^\mu)( dx_5 -
A^{(5)}_\mu dx^\mu)~.}}
Then the $D=4$ action is (see, {{\it e.g.}} \MaharanaMY)
\eqn\zb{S_4 = {1 \over 16\pi G_4} \int \! d^4 x \sqrt{-g}\,
e^{-2\phi_4} \left\{ R + 4(\p \phi_4)^2 +{1 \over 4} \p_\mu
G_{\alpha \beta} \p^\mu G^{\alpha \beta} -{1 \over 4 }G_{\alpha
\beta} F_{\mu\nu}^{(\alpha)} F^{(\beta)\mu\nu} \right\}~,}
where the indices $\alpha$ and $\beta$ run over $\psi$ and $z_5$, and $R$ is the
Ricci scalar of $ds_4^2$. The $D=4$ dilaton is
\eqn\zc{ e^{-2\phi_4} = \sqrt{\det G}e^{-2\phi_6} = \sqrt{\det
G}~.}
Also, the $D=4$ Einstein metric is
\eqn\zd{ g^{E}_{\mu\nu} = e^{-2\phi_4} g_{\mu \nu} = \sqrt{\det G}
~g_{\mu\nu}~.}
%


Of most interest are the asymptotic formulas for the four
dimensional quantities.  At $r=\infty$ the $T^2$ metric is
\eqn\uf{ G_{\alpha\beta}dx^\alpha dx^\beta = 4 Q_1 Q_5 \tilde{Z}_K
\left( dz^2 - {\hat{s}_\psi \over \sqrt{Q_1 Q_5 \tilde{Z}_K} }dz
d\psi +{R_K^2 + \hat{s}_\psi^2 \over 4 Q_1 Q_5 \tilde{Z}_K } d\psi^2
\right)}
where we defined the angular variable $z = x_5/R_5$.
$\hat{s}_\psi$ denotes the asymptotic value  following  from \de:
\eqn\ufa{\hat{s}_\psi = s_\psi |_{r=\infty} = - {2 \sqrt{Q_1 Q_5
\tilde{Z}_K} \over N_K \tilde{Z}_K}~.}
The $T^2$ metric corresponds to a torus with modular parameter
\eqn\ug{\tau ={1 \over 2 \sqrt{Q_1 Q_5 \tilde{Z}_K} }\left(
-\hat{s}_\psi +iR_K\right)~. }
By doing coordinate transformations preserving the periodicities
we can transform $\tau$ by $SL(2,Z)$.  However, the above $\tau$
depends on continuous moduli, and so we can't generically
transform it to a purely imaginary $\tau$.  In other words, we
can't transform away the  mixed $dz d\psi$ terms in the metric
\uf.

The asymptotic  string frame metric is
\eqn\ui{\eqalign{ d\hat{s}^2
&\approx -\left(1 - {Q_1+ Q_5 \over 2r}\right)dt^2 -{2 Q_1 Q_K Q_K
\over n R_5} {\sin^2 \theta \over r} dt d\phi +dr^2 + r^2
d\theta^2 + r^2 \sin^2 \theta d\phi^2~.  }}

The $D=4$ dilaton is
\eqn\uj{ e^{-2\phi_4} = \sqrt{\det G}= {1 \over \sqrt{Z_K}}
\approx 1 - {Q_K \over 2r}~. }

To read off the mass and angular momentum we need the following
two components of the asymptotic Einstein metric:
\eqn\uk{\eqalign{ g^E_{tt} &\approx -\left(1 - {Q_1+ Q_5 +Q_K
\over 2r}\right) \cr g^E_{t\phi} & \approx   -{Q_1 Q_5 Q_K \over n
R_5} {\sin^2 \theta \over r}~. }}

An asymptotically flat $D=4$ metric has the terms
\eqn\up{\eqalign{ g^E_{tt} &\approx  -( 1 - {2 G_4 M \over r})
\cr g^E_{t\phi} & \approx   -2G_4 J {\sin^2 \theta \over r} ~.}}
We therefore read off the mass and  angular momentum as
\eqn\uq{\eqalign {M & = {Q_1 +Q_5 +Q_K \over 4 G_4} \cr  J &={Q_1
Q_5 Q_K \over 2 n R_5 G_4}~. }}

The $D=4$ Newton constant is
\eqn\uo{ G_4 = {G_{10} \over V_6}={1 \over 8} {(2\pi)^4 g^2 \ap^4
\over R_K R_5 V_4},}
where we used
\eqn\uoa{  V_6 = (2\pi R_K)(2\pi R_5)
 V_4, \quad G_{10} ={1 \over 8}(2\pi)^6 g^2 \ap^4~. }
%
Also,
\eqn\uq{ Q_1 Q_5 =  {(2\pi)^4 g^2 \ap^4 \over 4 R_K^2 V_4 } N_1
N_5~.}
This then gives
\eqn\ur{ J=  \half { N_1 N_5 N_K\over n}~.}
This is precisely what we expect from the CFT point of view.  The
solutions considered above have $J>0$, but we can trivially get the
solutions with $J<0$ by time reversal.

We now work out  the gauge charges. The asymptotic gauge fields
are
\eqn\uh{ \eqalign{ 
A^{(\psi)}_\phi & = -Q_K \cos \theta + O({1 \over r}) \cr 
A^{(\psi)}_t & =-{1 \over R_K} \hat{k}_\psi =  {2 R_K Q_1
Q_5\over n R_5} {1 \over r} +O({1 \over r^2}) \cr 
A^{(5)}_\phi &= \hat{s}_\phi -{Q_K \over R_K} \cos \theta \hat{s}_\psi 
= O({1 \over r})  \cr 
A^{(5)}_t & = {\hat{s}_\psi \hat{k}_\psi \over R_K^2} = {
Q_1 Q_5 \over Q_K \tilde{Z}_K}{1 \over r} +O({1 \over r^2})~.}}
We need to take into account that in \zb\ the gauge fields mix via
$G_{\alpha \beta}$. To read off the charges we write the magnetic
potentials with upper indices, and electric ones with lower indices
(since the electric field corresponds to a canonical momentum.)

We immediately read off that $A^{(\psi)}$ has magnetic charge
$Q_K$, and corresponding quantized charge $N_m = N_K$. $A^{(5)}$
has vanishing magnetic charge.

 The electric potentials are then
\eqn\us{ \eqalign{ A_{(\psi) t} &= G_{\psi\psi} A^{(\psi)}_{t} +
G_{\psi 5} A^{(5)}_{t}  =(1+{s_\psi^2 \over R_K^2}) A^{(\psi)}_{t}
- {s_\psi \over R_K}  A^{(5)}_{t} = A^{(\psi)}_{t} \cr A_{(5) t}
&=G_{55} A^{(5)}_{t} + G_{5 \psi} A^{(\psi)}_{t}=A^{(5)}_{t}-
{s_\psi \over R_K} A^{(\psi)}_{t} =0 ~.}}
Therefore, the electric charge with respect to  $A_{(\psi)}$ is
nonvanishing, while it is vanishing for $A_{(5)}$. Next, we use the fact
that  $N_e$ units of quantized electric charge gives rise to the
long range potential
\eqn\usa{ A_{(\psi)t} = (16 \pi G_4) {N_e \over 4 \pi r},}
where we took into account the normalization factor of ${1 \over
16\pi G_4}$ in \zb.   We therefore read off
\eqn\ust{N_e = {R_K Q_1 Q_5 \over 2 n R_5 G_4}  = {R_K\over Q_K} J
= {2 J \over N_K} = N_1 N_5~.}
We find the relation
\eqn\usu{J= \half N_e N_m~.}
  As mentioned previously, this is the same angular momentum as
arises in ordinary electromagnetism from separated electric and
magnetic charges.

\subsec{Features of the singularity resolution}

As we have seen, the solutions containing a D1-D5-KK supertube wrapped
$n$ times have
\eqn\RRnn{ R_5 =   {2 \sqrt{Q_1 Q_5 \left(1 + {Q_K \over \Rt}\right)
}\over n}~. }

One can think of this relation as determining $\Rt$ in terms of
$R_5$; {\it i.e.} the separation between the ``location'' of the KK
monopole and  the location of the D1 and D5 branes as a function of
the compactification radius. This formula is analogous to the radius
formula for supertubes in flat space. Another feature similar to the
flat space case is that the ``radius'' $\Rt$ of the nonsingular
configuration decreases with increased dipole charge.

A more unexpected feature of  equation \RRnn\ is that as $R_5 $
approaches $ {2 \sqrt{Q_1 Q_5 } \over n}$ we find that $\Rt$  goes
to infinity. Hence, for fixed  charges it is possible to change the
moduli of the solution only within some range.  Although this
behavior is perhaps unexpected for the asymptotically flat geometry,
if it persisted in the near horizon limit it would be truly
peculiar, with no obvious CFT interpretation. Fortunately, after
taking the near-horizon limit, formula \RRnn\ becomes
\eqn\RRnh{ R_5 = {2\sqrt{Q_1 Q_5 } \sqrt{Q_K \over \Rt }\over n} }
and there is no longer a lower bound on $R_5$.

Physically, the reason why the supertube disappears from the spectrum
for sufficiently small $R_5$  is that the space at infinity is not
$\IR^4$ but $\IR^3 \times S^1$. If we think about Taub-NUT space as
a cigar, then small supertubes sit near the tip of the cigar.  As
the supertube radius increases (this can be done by changing
moduli), the tubes become larger and slide away from the tip, while
wrapping the cigar. Since the radius of the cigar is finite, the
tubes will eventually slide off to infinity and disappear.

It is interesting to note that although the D1-branes and
D5-branes are smeared along the KKM fibre in both the naive and
correct geometries, at $r=0$ this fibre
shrinks to zero size. Therefore, the fact that the D1 and D5 branes move away
from the origin and acquire a KKM dipole moment is indeed an expansion
into a supertube.
However, from the 4D point of view this expansion is not easily seen, since
both the unexpanded branes and the supertube reduce to a point when
compactified to four dimensions.

Hence, from a four-dimensional point of view our solutions contain
just two sets of charges separated by a certain distance. If one
tries to take this separation to zero the solution is the naive
singular geometry \bc. Hence, from a four-dimensional perspective
the singularity of the naive geometry is resolved by the splitting
of the brane sources. Moreover, for certain values of the separation
the resulting configuration is a bound state with a clear CFT dual
description, while for other values it is not.

We see that not any splitting of the branes into distinct stacks
will resolve the singularity, but only a special type of split (with
the KK-monopoles  in one stack and the D1 and D5 branes in the
other).  From a supergravity perspective it is not always clear when
a given solution is physically acceptable or not; it depends on the
duality frame chosen.   The IIB duality frame employed here admits
manifestly smooth geometries, which are thus physically allowed. But
in other duality frames these geometries will be singular, and one
needs other criteria to determine their physical relevance.  One
such method is to give an open string description of the
corresponding object, as can be done for the original supertube
$D(p-2) + F1 \rightarrow Dp$.  Then the microscopic description will
yield the necessary constraints on the splitting.

\newsec{Discussion}

In this paper we found asymptotically flat solutions representing
the $U(1) \times U(1)$ invariant chiral primaries of the D1-D5-KK
system.  We found that these solutions are either completely
smooth or have acceptable orbifold singularities due to coincident
KK-monopoles.  These solutions have several novel features.
One is the separation of the D1-D5 charges from
that of the KK-monopole, in the sense that the corresponding
harmonic functions are sourced at different locations in $\IR^3$.
Another feature is that the solution carries an
electric charge with respect to the same gauge field that is
magnetically charged.  The charges combined together to obey the
relation $J= \half N_e N_m$, which also appears in pure
electromagnetism.

It would clearly be desirable to relax the condition of
$U(1)\times U(1)$ symmetry so as to be able to find the full set
of chiral primaries.  It is likely that the corresponding
supergravity solutions will have the Taub-NUT metric replaced by a
less symmetric hyper-Kahler manifold, since there would be no
obvious reason for the four-dimensional base metric to preserve
more symmetry than the full solution.

By a chain of dualities we can transform our D1-D5-KK solution into
one carrying charges D1-D5-P, corresponding to the canonical
five-dimensional black hole.   Therefore, it is appropriate to ask
to what extent our solutions can be thought of as the microstates of
the D1-D5-P system.   The main issue is that our solution is
asymptotically $\IR^{(3,1)}\times T^6$, while the finite entropy
black hole of the D1-D5-P system is asymptotically
$\IR^{(4,1)}\times T^5$.  If we perform duality transformations at
the level of supergravity there is no possibility of transforming
between these two types of solutions.  The dualities would instead
produce a D1-D5-P system smeared over a transverse circle.

\lref\GregoryTE{
  R.~Gregory, J.~A.~Harvey and G.~W.~Moore,
  ``Unwinding strings and T-duality of Kaluza-Klein and H-monopoles,''
  Adv.\ Theor.\ Math.\ Phys.\  {\bf 1}, 283 (1997)
  [arXiv:hep-th/9708086].
}

Nevertheless, it is possible that a more accurate duality
transformation would avoid this problem.  The key step in the
duality chain is the T-duality along $\psi$, the fibre direction
of the KK-monopole.  The D1-branes and D5-branes are both
delocalized in this direction, but let us ignore them for the
moment, so that we are just considering the T-duality of a
KK-monopole.  The T-duality produces an NS5-brane, and the
question is whether this NS5-brane is smeared or localized over
the dual circle.  The standard Buscher rules \Buscher\ certainly
produce a smeared solution. However, as discussed in \GregoryTE\
and shown explicitly in \tong, an exact CFT treatment of the
T-duality yields a localized NS5-brane.  One way to see this is
that the CFT derivation of T-duality involves gauging the
translational isometry of the original circle. This requires
introducing a corresponding $U(1)$ gauge field, which is
subsequently integrated out.  The point is that there are
instantons in this gauge field which violate the translational
isometry of the dual circle.

We might expect a similar phenomenon to occur in our case, leading
to a solution which is asymptotic to the standard D1-D5-P
solution. Of course, since we have a much more complicated setup
than just a KK-monopole, involving RR-fields and the 1-forms $k$
and $s$,  it is hard to give a direct argument for this. One
indirect way to see that this phenomenon is likely to occur in our
case is to recall that T-duality interchanges winding and momentum
modes. Since  winding number is not conserved in our backgrounds
(due to the contractibility of the $S^1$ fibre of Taub-NUT), the
resulting T-dual background will not preserve momentum, and hence
it will not be a smeared collection of branes, but a localized
one.
 While it
would certainly be desirable to directly write down solutions for
the microstates of D1-D5-P, if the above reasoning is correct we
may at least be able to extract some of the physics of these
microstates by studying our dual D1-D5-KK solutions.

As we have discussed, our solutions have a clear microscopic meaning
in the D1-D5-KK CFT.  They are the chiral primaries, or
equivalently, the Ramond ground states.  In the effective string
language, these are described by $N_1 N_5 N_K / n$ effective strings,
each of length $n$.  On the other hand, the standard finite entropy
black hole of the D1-D5-KK-P system corresponds to taking a single
effective string of length $N_1 N_5 N_K$ and adding momentum to it.
To preserve susy, the momentum is added to the non-supersymmetric
side of the $(4,0)$ CFT; the excitations carry no R-charge, and
hence the black hole carries no angular momentum. It is natural to
consider combining these two elements.  That is, to consider
dividing the full effective string into two parts, one of which
is a collection of short effective strings in the Ramond
ground state, and the other is a single effective string carrying
momentum.   In \BenaTK, this sort of configuration in the D1-D5 CFT
was argued to correspond to five-dimensional BPS black rings.  With
this in mind, we expect that these configurations in the D1-D5-KK
CFT will correspond to the solutions we have found in this paper,
except that the two-charge supertube is replaced by the three charge
BPS black ring.  That is, we would have a BPS black ring whose ring
direction is wrapped around the nontrivial $S^1$ of Taub-NUT.  From
the four-dimensional point of view this would be a black hole, since
the ring extends along a compact direction.  So we are led to the
prediction of a new four-dimensional BPS black hole solution
carrying nonzero angular momentum.

Such a solution could also have been anticipated in another way.
In \BenaTK\ it was noted that the charges of the five-dimensional
black ring correspond to the charges appearing in the quartic
$E_{7(7)}$ invariant.  However, one of the charges was actually
vanishing for the black rings discussed there, and it was noted
that it corresponded to a KK-monopole charge.  Now we see that
this missing charge is precisely that of the KK-monopole discussed
in this paper.  The new solution we are conjecturing will combine
all these charges together.

\bigskip
\noindent {\bf Acknowledgments:} \medskip \noindent We would like
to thank Don Marolf and Nick Warner for discussions. The work of IB
and PK is supported in part by  NSF grant  PHY-00-99590.

\newsec{Appendix: Singularity analysis}

In this Appendix we analyze the potential singularities in
\eqn\appa{ ds_{10}^2= {1 \over \sqrt{Z_1 Z_5}}\left[ -(dt+k)^2
+(dx_5 -k -s)^2 \right]+ \sqrt{Z_1 Z_5} ds_{K}^2+\sqrt{{Z_1 \over
Z_5}}ds_{T^4}^2~}
with
\eqn\appb{Z_{1,5}= 1 +{Q_{1,5} \over \Sigma},\quad \Sigma
=\sqrt{r^2 +\Rt^2 +2 \Rt r \cos \theta}~,}
\eqn\df{ds_K^2  =Z_K(dr^2 + r^2 d\theta^2 + r^2 \sin^2 \theta
d\phi^2)+{1 \over Z_K}\left(R_K d\psi + Q_K \cos \theta
d\phi\right)^2~,}
and with the 1-forms $k$ and $s$ given in \rb, \cd, and \rmm.

For generic choices of parameters $c_i$, $d_i$ and ring radius
$\Rt$, our solution will have curvature singularities at $r=0$ and
$\Sigma=0$, and Dirac-Misner string singularities at $\sin
\theta=0$.  In this appendix we show that all the free parameters
of the solution are fixed by demanding smoothness.  We will then
generalize to allow for $\ZZ_n$ singularities corresponding to $n$
coincident KK-monopole supertubes.

In the following we suppress the trivial $T^4$ part of the metric,
since it is manifestly nonsingular.

\subsec{ $r=0$ singularities}

Viewing $k$ as a 1-form on the Taub-NUT metric, we demand that $k$
is nonsingular at $r=0$.  Otherwise there will be singular terms
in the metric of the form $dt d\psi$ and $dt d\phi$.  Since the
angular coordinates break down at $r=0$, finiteness of $k$
requires that $k_\psi$ and $k_\phi$ vanish at $r=0$.

 We find the leading small $r$ behavior
\eqn\rr{\eqalign{k_\psi &\approx -{c_2 R_K\over r} -c_1R_K -
{c_3R_K \over \Rt}+{d_2 R_K\over 2 Q_K} \cr  k_\phi &\approx
{Q_K\over R_K} \cos \theta k_\psi+{Q_K \over \Rt}c_3+ c_4-\half
d_4+(-\half d_3 -c_3 +Q_K c_1 -c_2 -\half d_2)\cos\theta}}
and so we demand the following four conditions
\eqn\rs{ c_2 =0, \quad {Q_K \over \Rt}c_3+ c_4-\half d_4=0, \quad
- \half d_3 -c_3 +Q_K c_1  - \half d_2=0,\quad c_1+{1 \over
\Rt}c_3 - {1 \over 2Q_K} d_2=0~. }

Next, we focus on  the small $r$ behavior of $s$, which is
\eqn\rv{s \approx \bar s_{\psi} d\psi +
\bar s_{\phi} d\phi~.}
 where $\bar s_{\psi}= (d_5-d_2 {R_K \over Q_K}) $ and
$  \bar s_{\phi} = (d_3 + d_4) $. If we
were to demand that $s$ be a well defined 1-form on Taub-NUT we
would require $\bar s_\psi = \bar s_\phi = 0$, as for $k$.
However, taking into
account the nontrivial mixing of the angular Taub-NUT coordinates
with $x_5$, we can in fact relax this condition and still obtain a
nonsingular metric.  This is most easily seen by transforming to
the following new coordinates:
\eqn\rwa{\eqalign{r&= {1 \over 4 }\rh^2 \cr \theta &=2\thetah \cr
x_5 & = R_5 \hat \gamma \cr \phi& = \phih-\psih-{1 \over R_5}x_5
\cr \psi& = {Q_K\over R_K} (\psih+\phih-{1 \over R_5}x_5)~.}}
$\hat{\gamma}$ is $2\pi$ periodic, while $(\psih,\phih)$ have
periodicities
\eqn\rwab{(\psih,\phih) \cong  (\psih,\phih+2\pi )  \cong
(\psih+{2\pi \over N_K},\phih+{2\pi \over N_K})~.}
   Assuming \rs, and
thereby solving for $c_1, c_2, d_3, d_4$, and also writing
\eqn\rw{\eqalign{d_2 &= {Q_K \over R_K} d_5 +\half R_5 , \quad c_4
= c_3 -{1 \over 4} R_5 }}
the leading behavior of the metric is
\eqn\rx{\eqalign{ds^2 &\approx -{\Rt \over \sqrt{Q_1 Q_5}}dt^2
+{R_5^2 \Rt \over \sqrt{Q_1 Q_5}}d\phih^2 \cr &\quad + {\sqrt{Q_1
Q_5}Q_K\over \Rt} \left\{ d\rh ^2+\rh^2  (d\thetah^2 +\sin^2
\thetah  d\psih^2 +\cos^2 \thetah d\hat \gamma^2 )\right\}~.}}
This metric is  smooth given the identification in \rwab.  Note in
particular that the $\ZZ_{N_K}$ identification includes a shift of
the fixed size $\phih$ circle; therefore there are no fixed points.

Besides the new coordinates \rwa , there are other coordinate
transformations that give smooth metrics. The first two
coordinate changes are the same as in \rwa, while the last three
are modified.
%
%
We choose one of the angles to be proportional to the combination
of $x^5,\psi$, and $\phi$ that appears in the vielbein containing $dx^5$:
\eqn\efive{{1 \over R_5} dx_5  - {\bar s_{\psi} \over R_5} d \psi -
{\bar s_{\phi} \over R_5 } d \phi \equiv {1 \over R_5} dx_5 - n_1 d
\psi + \left(n_1 {N_K \over 2}- n_2\right) d \phi \equiv d \hat
\phi~. }
If we further define
\eqn\psihat{\psih = {\psi \over N_K} - {\phi \over 2}~,~~~~ 
\hat \gamma = \left( { n_1 \over n_2}-{1 \over N_K}  \right)
\left(\psi - {N_K \over 2} \phi \right)  -
  {x^5 \over n_2R_5 }}
then the metric becomes again \rx, except that the identifications
of the coordinates are now:
\eqn\ident{\eqalign{\left(\phih, \psih, \hat \gamma\right) & \cong
\left(\phih + 2 \pi n_1, \psih + {2 \pi \over N_K}, \hat \gamma + {2
\pi}\left(  { n_1 \over n_2} - {1 \over N_K} \right) \right)\cong
\cr &\cong \left(\phih+2 \pi, \psih, \hat \gamma -
 {2 \pi \over n_2}\right) \cong \left(\phih + 2
\pi n_2, \psih, \hat \gamma\right)~. }}

Of course, in order for these identifications to give a smooth space,
both $n_1$ and $n_2$ must be
rational. One can also see that the choice of constants in \rw\ and
\rwa\ is equivalent to taking
\eqn\gauge{n_2 = -1~,~~~~~ n_1 = - {1 \over N_K}~.}

It is interesting to explore the physical meaning of the extra
parameters $n_1$ and $n_2$. Their only effect is to add two
constants to $s_{\psi}$ and $s_{\phi}$ in equation \dez. Adding a
constant to $s_{\phi}$ is a trivial diffeomorphism transformation.
Adding a constant to $s_{\psi}$ changes the mixing of $d \psi$ and
$d x^5$ at infinity, and affects the modular parameter of the torus
that one uses to obtain the four-dimensional theory.

The new value of $s_{\psi}$ at infinity is \eqn\spsiinf{\hat s_\psi
= n_1 R_5 + R_K \sqrt{Q_1 Q_5 \over \tilde R ( Q_K +\tilde R  )} ~,}
and for some values of the moduli it is possible to find a rational
$n_1$ that gives $\hat s_\psi = 0$. However, even if the torus is
diagonal, the gauge potentials that appear in equations \uh\ and
\us\ are modified, and in particular the four-dimensional
KK-monopole charge does not point in the $\psi$ direction but in a
combination of $x_5$ and $\psi$. The gauge choice \gauge\ aligns the
KK-monopole charge and the electric charge along the $\psi$
direction, and we will be choosing it from now on.

\subsec{Dirac-Misner string singularities}

As written in \df, the Taub-NUT metric has coordinate
singularities at $\sin \theta =0$.  These can be removed by
shifting $\psi$.  In particular, at $\cos \theta = \pm 1$ the
metric involves the combination $ d\psi  \pm \half N_K d\phi$, and
so the shift is $\psi \rightarrow \psi \mp \half N_K \phi$.

If the 1-forms $k$ and $s$ are proportional to $ d\psi  \pm \half
N_K d\phi$ at $\cos \theta = \pm 1 $, then the shift $\psi
\rightarrow \psi \mp \half N_K \phi$ will remove the offending
$\phi$ components.  Using \rs\ and \rw, it is straightforward to
verify that at $\cos \theta =1$ both $k$ and $s$ are in fact
proportional to $ d\psi + \half N_K d\phi$.   At $\cos \theta =-1$
we find that $k$ is proportional to $ d\psi  - \half N_K d\phi$,
as is $s$ in the region $r> \Rt$.   But for $r<\Rt$ this does not
hold, and the situation is more involved.  In particular,  for
$\cos \theta =-1$ and $r<\Rt$ we find: 
\eqn\tta{\eqalign{ k& = -2 k_\phi d\psih, 
\quad k+s =  -2(k_\phi+s_\phi)d\psih -
 R_5 (d\psih+d\phih)+dx_5~, }}
with angular coordinates defined as in \rwa. We then note that
$dx_5$ appears in the metric via the combination $dx_5-k-s$.  This
indicates that the contribution to $g_{55}$ from this term
vanishes. Moreover, as one can see from \rwa\ and \df, the
contribution to $g_{55}$ from the Taub-NUT vanishes at $\cos
\theta = -1$. In the hatted coordinates system and near
$\cos\theta =-1$, $dx_5$ and $d\theta$ appear in the combination
$d(\pi -\theta)^2+ (\pi-\theta)^2 {1 \over R_5^2} dx_5^2$, which
shows that the $x_5$ circle smoothly shrinks to zero size.  On the
the other hand, the $\psih$ and $\phih$ circles stabilize at
finite size.  Thus the complete metric is smooth at $\cos \theta
=-1$ and $r<\Rt$.

\subsec{Checking the metric and forms at $\Sigma=0$. }

The periodicities of the coordinates $\psi,\phi,x_5$ appearing in \df\ are:
\eqn\period{(\psi,\phi,x_5 ) \cong (\psi,\phi, x_5+ 2 \pi R_5 )
\cong (\psi + N \pi , \phi + 2 \pi, x_5 ) \cong ( \psi + 2 \pi ,
\phi, x_5 )~. }

In order to check the behavior of the metric and forms at the point
$\Sigma = 0$ it is good to transform to a coordinate system in which
this point is the origin of  the $\IR^3$ that forms the base of the
Taub-NUT space \df:
\eqn\dfs{ds_K^2  =Z_K(d\Sigma^2 + \Sigma^2 d\theta_1^2 + \Sigma^2 \sin^2 \theta_1
d\phi^2)+{1 \over Z_K}\left(R_K d\psi + Q_K \cos \theta d\phi\right)^2~,}
where $\Sigma $ is defined in \appb, $\cos \theta_1 = {\Rt + r \cos
\theta \over \Sigma} $, and $\phi $ is  unchanged After substituting
$s$ and $k$ in the metric, and expanding the metric components
around $\Sigma=0$, the  leading components of the metric are
nondiagonal. Moreover, the leading components of the metric in the
$\psi\psi$, $ \phi \phi$ and $ \psi \phi$ directions blow up like
\eqn\gpsi{ {Q_1 Q_5 - 4 c_3^2 \tilde{Z}_K \over \Sigma}~.} To render
these components finite we then must choose
\eqn\tg{ c_3 = \half \sqrt{Q_1 Q_5 \over \tilde{Z}_K}~.}
After making this substitution one can make the leading metric diagonal by
making the  transformation:
\eqn\dia{t = l_0~ \bar t ~,~~~ x_5 = R_5 \eta - l_0 \bar t ~,~~~
\phi = \phi~,~~~\psi = {N_K \over 2} (- \bar t  -  \gamma + \phi)}
where $l_0$ is a finite constant\foot{For the curious,
\eqn\lzero{
{l_0 \equiv \frac{Q_K\,\left( Q_K\,{Q_1}\,{Q_5} + (Q_1+ Q_5)\,\Rt (Q_K + \Rt) \right)
}{{\sqrt{{Q_1Q_5 \Rt}}}\, \,{\left( Q_K + \Rt \right) }^{\frac{3}{2}}}}.}}.
From \period\ and \dia\ one can see that the period of $\phi $ and $\eta$ is $2 \pi$,
while the period of  $\gamma$ is $4 \pi \over N$.


After the diagonalization, the components of the metric $g_{tt}$ 
and $g_{\gamma \gamma}$
are finite, while  the leading metric in the $\Sigma, \theta_1, \eta, \phi$ 
directions is
\eqn\lmet{\eqalign{& \sqrt{Q_1 Q_5}~{\tilde{Z}_K } \times \cr
&\left[{d \Sigma^2 \over \Sigma} + \Sigma (d \theta_1^2 + \sin^2
\theta_1 d \phi^2) +
 { R_5^2 ~ \Sigma \over Q_1 Q_5 \tilde{Z}_K}
 \left(d \eta + {\sqrt{Q_1 Q_5 \tilde{Z}_K } \over R_5} (1+ \cos \theta_1)
 d \phi \right)^2 \right]~.  }}
It is clear that when
\eqn\RR{ R_5 = 2\sqrt{Q_1 Q_5 \tilde{Z}_K } }
this metric becomes the metric at the origin of a new Taub-NUT
space, in which $\Sigma$ plays the role of a radius and $d \eta$
that of a fiber. The fact that the metric near the location of the
D1 branes and D5 branes can be written as a Taub-NUT space indicates
that the D1 branes and D5 branes have formed a D1-D5-KK supertube at
that location. Hence our solution contains a D1-D5-KK  supertube at
$\Sigma = 0 $ and a KK-monopole at $r=0$. When \RR\ is satisfied,
the dipole charge of the KK-monopole of the supertube is 1, and the metric
\lmet\ is manifestly smooth.

We can also consider metrics in which the KKM supertube is wrapped $n$
times. These metrics have
a $\ZZ_n$ orbifold singularity at $\Sigma = 0$, and are obtained by
simplify modifying \RR\ to:
\eqn\RRn{ R_5 = {2 \sqrt{Q_1 Q_5 \tilde{Z}_K } \over n}~.  }
From the point of view of string theory, these $\ZZ_n$ orbifolds are nonsingular,
and so we should allow them in our class of solutions.

We now turn to checking the smoothness of the RR 2-form potential of
our solutions. The only place where the RR fields might be divergent
is at $\Sigma = 0$, where the D1 branes and the D5 branes are located.
Both the smoothness of the metric, and the fact that the near-horizon
region of these solutions can be mapped to AdS$_3$ in global
coordinates clearly point to the absence of any divergences. However,
it is instructive to see how this happens.

As we have discussed in Section 3, the ``electric'' RR potential of the solution is:
\eqn\rre{C^2_{e} = Z_1^{-1} (dt + k_\psi d\psi + k_\phi d \phi) \wedge (dx^5 - s_\psi d\psi - s_\phi d \phi - k_\psi d\psi - k_\phi d \phi )  }
while the magnetic one is given implicitly by
\eqn\rrm{d C^2_{m} = - \star_4 (d Z_5)   }
where $\star_4$ is the Hodge dual on the Taub-Nut metric \df.
After making the coordinate change \dia,  the leading components of the
electric potential near the point $\Sigma = 0$ are:
\eqn\rres{ C^2_{e} \sim \Sigma \left( d \eta + {1 + \cos \theta_1 \over 2} d \phi
+ {c_1 \over \Sigma}(d \gamma + d \bar t ) \right) \wedge (d \bar t
+ {c_2 \over \Sigma}(d \gamma + d \bar t ) )}
where $c_1$ and $c_2$ are constants that can be determined straightforwardly
from the metric. The part proportional to ${1 \over \Sigma}$ cancels, and most
of the constant forms are not dangerous because the angles they contain never
shrink to zero size. The only possibly dangerous component of $C^2_e$ is
\eqn\cedang{ (d \gamma + d \bar t ) \wedge  \left( d \eta + {1 +
\cos \theta_1 \over 2} d \phi \right)~. }
However, since $g_{ \bar t \bar t }$ and $g_{\gamma \gamma  }$ are
finite,  and the second parenthesis is nothing but the fibre of
the Taub-NUT space \lmet, this component is also benign.

Since the harmonic function $Z_5$ is very simple when written in
terms  of $\Sigma$, the magnetic field strength can be easily
evaluated to be:
\eqn\rrms{d C^2_{m} = R_K Q_5 \sin \theta_1  d\theta_1\wedge
d \phi \wedge d \psi  }
and hence one can write the potential that gives rise to this field strength
as
\eqn\rrmsp{ C^2_{m} = Q_K Q_5 \left(  2 d \eta + (1 + \cos \theta_1)  d \phi
\right) \wedge (d \gamma + d \bar t  )  }
which is again proportional to the fibre of the Taub-NUT space,
and  hence regular.

We have therefore verified that with suitable choice of parameters
the metric is everywhere smooth. Solving for the parameters yields
the values in \daa.

\listrefs
\end